\documentclass[pra,twocolumn,preprintnumbers,superscriptaddress]{revtex4-2} 

\usepackage{graphicx}
\usepackage{bm}
\usepackage{amsmath}
\usepackage{amssymb}
\usepackage{amsfonts}
\usepackage{fancyhdr}
\usepackage{slashed}
\usepackage{latexsym,epsfig,bbm}

\newcommand{\bra}[1]{\langle{#1}|}
\newcommand{\ket}[1]{|{#1}\rangle}

\newcommand{\braket}[2]{\langle{#1}|{#2}\rangle}
\newcommand{\bopk}[3]{\langle{#1}|{#2}|{#3}\rangle}

\newcommand{\figref}[1]{Fig.~\ref{#1}}

\usepackage{color}
\definecolor{blue}{rgb}{0,0.2,1}

\definecolor{red}{rgb}{0.9,0,0}

\newcommand{\kb}{k_{\mathrm{B}}}

\begin{document}

\title{Memory compression and thermal efficiency of quantum implementations\\ of non-deterministic hidden Markov models}

\author{Thomas J.~Elliott}
\email{physics@tjelliott.net}
\affiliation{Department of Mathematics, Imperial College London, London SW7 2AZ, United Kingdom}

\date{\today}

\begin{abstract}
Stochastic modelling is an essential component of the quantitative sciences, with hidden Markov models (HMMs) often playing a central role. Concurrently, the rise of quantum technologies promises a host of advantages in computational problems, typically in terms of the scaling of requisite resources such as time and memory. HMMs are no exception to this, with recent results highlighting quantum implementations of deterministic HMMs exhibiting superior memory and thermal efficiency relative to their classical counterparts. In many contexts however, non-deterministic HMMs are viable alternatives; compared to them the advantages of current quantum implementations do not always hold. Here, we provide a systematic prescription for constructing quantum implementations of non-deterministic HMMs that re-establish the quantum advantages against this broader class. Crucially, we show that whenever the classical implementation suffers from thermal dissipation due to its need to process information in a time-local manner, our quantum implementations will both mitigate some of this dissipation, and achieve an advantage in memory compression.
\end{abstract}
\maketitle 

\section{Introduction}

\emph{Hidden Markov models} (HMMs) provide a powerful representative tool for modelling stochastic systems~\cite{rabiner1986introduction}. They are able to generate a diverse range of complex, non-Markovian stochastic processes, finding applications in a broad spectrum of fields, including speech recognition~\cite{rabiner1989tutorial}, time-series analysis~\cite{crutchfield1994calculi, yang2020measures}, cryptanalysis~\cite{karlof2003hidden}, machine learning~\cite{ghahramani1996factorial, fine1998hierarchical, seymore1999learning}, bioinformatics~\cite{baldi1994hidden, krogh2001predicting, stanke2003gene}, economics~\cite{bhar2004hidden}, and statistical physics~\cite{gammelmark2014hidden}. Given their near-universal deployment across the quantitative sciences, they form an essential topic of research in their own right.

Akin to how quantum mechanics revolutionised physics in the early 20th century, quantum information processing promises to do the same for computational science~\cite{nielsen2000quantum}. Technologies based on this paradigm offer a means to implement better, faster, more efficient algorithms and protocols~\cite{montanaro2016quantum}. Naturally, this has spurred investigations into quantum extensions of HMMs, including characterisation of their expressivity~\cite{wiesner2008computation, monras2011hidden, o2012hidden, clark2015hidden, cholewa2017quantum, adhikary2020expressiveness}, how they can be inferred~\cite{monras2016quantum, srinivasan2018learning, ho2020robust}, and how they can outperform classical automata~\cite{gruska2015potential, tian2019experimental}.

Two key areas where quantum HMMs manifest advantages are in stochastic simulation and information engines. \emph{Quantum implementations} of deterministic HMMs offer memory compression in the former~\cite{gu2012quantum, mahoney2016occam, aghamohammadi2018extreme, elliott2018superior, binder2018practical, elliott2019memory, liu2019optimal, loomis2019strong}, and greater thermal efficiency in the latter~\cite{loomis2020thermal}, compared to corresponding classical implementations. The restriction of determinism is sometimes a necessity:
non-deterministic consumers of information reservoirs have been shown to be capable of violating the second law of information thermodynamics~\cite{garner2019oracular}; and adaptive systems must typically act in a causal manner~\cite{barnett2015computational, thompson2017using}. In many other contexts however, the broader class of non-deterministic HMMs can be employed, against which the current quantum advantages do not hold. Here, we generalise quantum implementations to encompass them and recover the advantages.

We introduce a systematic approach to constructing quantum implementations of non-deterministic HMMs, showing that memory and thermodynamical advantages arise for almost any HMM. Our findings strengthen, extend, and supersede analogous prior results~\cite{liu2019optimal, loomis2020thermal, loomis2020thermodynamically}, which were limited in scope by considering an incomplete suite of possible quantum implementations.

\section{Processes, generators, and implementations}

A bi-infinite, discrete-time, discrete-event stochastic process~\cite{khintchine1934korrelationstheorie} is characterised by a sequence of random variables $X_t$ taking on values $x_t\in\mathcal{X}$, with the index labelling the timestep $t\in\mathbb{Z}$. Throughout, we will use upper-case to denote random variables, and lower-case the corresponding variates. A contiguous block of the sequence of length $L$ is denoted $X_{t:t+L}:=X_t\ldots X_{t+L-1}$; a process $\mathcal{P}$ is then defined by the distribution $P(X_{-\infty:\infty})$. We restrict our attention to stationary stochastic processes, wherein $P(X_{0:L})=P(X_{t:t+L})\forall L,t\in\mathbb{Z}$. We take the index $t=0$ to represent the present timestep, such that $X_{-\infty:0}$ is the past of the process, and $X_{0:\infty}$ its future.

We consider HMMs~\cite{upper1997theory} that give rise to such processes. We will here focus on edge-emitting (or Mealy) HMMs, while noting that our results can be readily adapted to state-emitting (Moore) HMMs~\cite{shalizi2001computational}. An edge-emitting HMM $(\mathcal{S},\mathcal{X},\{T_{s's}^x\})$ consists of a (potentially infinite) set of states $s\in\mathcal{S}$, an alphabet of symbols $\mathcal{X}$, and a transition structure $\{T^x_{s's}:=P(s',x|s)\}$ describing the probability of a system in state $s$ transitioning to state $s'$ while emitting symbol $x$. We deal with irreducible HMMs, such that any state $s'\in\mathcal{S}$ can be reached by any state $s\in\mathcal{S}$ with non-zero probability after a sufficient number of transitions; this can be diagnosed by verifying that the matrix $T_{s's}=\sum_xT_{s's}^x$ is irreducible. An irreducible HMM has a unique stationary state $\pi(s)$.

The emitted symbols of a HMM generate a stochastic process, described by the distribution
\begin{equation}
P(x_{0:L}):=\sum_{s_0\ldots s_{L}}T_{s_{L}s_{L-1}}^{x_{L-1}}\ldots T_{s_1s_0}^{x_0}\pi(s_0).
\end{equation}
We refer to the HMM as a \emph{generator} of the process. An important class of generators are for \emph{deterministic} (sometimes referred to as \emph{unifilar}) HMMs~\cite{shalizi2001computational}, for which the end state of a transition is uniquely determined by the start state and emitted symbol, i.e., for any given $s$ and $x$, $T^x_{s's}$ is non-zero for at most one $s'$. When the HMM is deterministic, we say that the generator is \emph{predictive}. The counterparts to this are \emph{co-unifilar} HMMs and \emph{retrodictive} generators, for which the start state is uniquely determined by the end state and emitted symbol~\cite{crutchfield2009time}.

With classical information processing a physical instantiation of a HMM must encode $\{s\}$ as distinguishable states $\{\ket{s}\}$ of a system, while with quantum information processing we can map them to (generally) non-orthogonal quantum states $\{s\}\to\{\ket{\sigma_s}\}$~\cite{liu2019optimal}.

{\bf Definition 1} (Quantum implementations): \emph{A quantum implementation $(\{\ket{\sigma_s}\},\Phi)$ of a generator $(\mathcal{S},\mathcal{X},\{T_{s's}^x\})$ is a set of quantum states $\{\ket{\sigma_s}\}$ and a quantum channel $\Phi$ which satisfy}
\begin{equation}
\label{eqchanneldef}
\Phi(\ket{\sigma_s}\bra{\sigma_s})=\sum_{s'x}T_{s's}^x\ket{\sigma_{s'}}\bra{\sigma_{s'}}\otimes\ket{x}\bra{x}.
\end{equation}

Being defined with respect to a generator~\cite{loomis2020thermal} this is a subtly different notion to that of a quantum model~\cite{gu2012quantum}, which is defined with respect to a process, though a quantum implementation also represents a quantum model of the process generated by the generator it implements.

\section{Memory compression} 

One way the performance of a generator and its implementation can be evaluated is in terms of the amount of memory it requires. Two key quantifiers of this are the size of the memory state space, and the amount of information it stores. With respect to a generator $g$, we denote~\cite{crutchfield1989inferring, ruebeck2018prediction,thompson2018causal,liu2019optimal}:
\begin{align}
\label{eqmeasures}
D_g:=&\log_2[\mathrm{rank}(\rho)];\nonumber\\
C_g:=&-\mathrm{Tr}(\rho\log_2[\rho]),
\end{align}
where $\rho=\sum_s\pi(s)\ket{\sigma_s}\bra{\sigma_s}$ is the stationary state of the memory. For classical implementations these reduce to the (logarithm of) the number of states and the Shannon entropy of the stationary distribution $\pi(s)$ respectively.

A particular privileged generator of a process is the $\varepsilon$-machine of computational mechanics~\cite{crutchfield1989inferring, shalizi2001computational, crutchfield2012between}, representing the predictive generator that can be classically-implemented with the minimal amount of memory according to both measures of Eq.~\eqref{eqmeasures}. The corresponding measures $D_\mu$ and $C_\mu$ are known as the topological and statistical complexity. Recently, a growing body of work has established that quantum implementations of $\varepsilon$-machines (with memory costs $D_q$ and $C_q$) can generally undercut this minimality~\cite{gu2012quantum, mahoney2016occam, aghamohammadi2018extreme, elliott2018superior, binder2018practical, elliott2019memory, liu2019optimal, loomis2019strong}: $D_q\leq D_\mu$ and $C_q\leq C_\mu$, with the compression advantage sometimes able to grow unboundedly large~\cite{aghamohammadi2017extreme, garner2017provably, elliott2018superior,elliott2019memory, thompson2018causal, elliott2020extreme, elliott2021quantum}. The present state-of-the-art quantum implementations~\cite{liu2019optimal} are defined implicitly through a unitary interaction:
\begin{equation}
\label{eqdeterministic}
U\ket{\sigma_s}\ket{0}=\sum_{x}\sqrt{P(x|s)}e^{i\varphi_{sx}}\ket{\sigma_{\lambda(s,x)}}\ket{x},
\end{equation}
where $\{\varphi_{sx}\in\mathbb{R}\}$ are free parameters (to be optimised over) and $\lambda(s,x)$ is the predictive update rule that determines the subsequent memory state. Decoherence of the output register in the computational basis (by, e.g., a non-selective measurement, or coupling to an external ancilla) results in a channel of the form of Eq.~\eqref{eqchanneldef}. However, this construction is valid only for predictive generators.

While there are many scenarios where one would require predictive behaviour, such a restriction is not necessary when one is purely concerned with generating a stochastic process. Pertinently, for some processes generators based on non-deterministic HMMs can be classically implemented less memory than the $\varepsilon$-machine~\cite{crutchfield1994calculi, lohr2009non, ruebeck2018prediction}: $D_g<D_\mu$ and $C_g<C_\mu$. It behooves us to ask whether analogous quantum implementations of non-predictive generators can be constructed, and whether they exhibit similar memory compression advantages. We now answer both these questions in the affirmative.

We begin by assuming that pre-processing has been performed on the classical generator to prune any redundancy due to states with equivalent future morphs. That is, if two states $s,s'\in\mathcal{S}$ satisfy $P(X_{0:\infty}|S_0=s)=P(X_{0:\infty}|S_0=s')$, then we merge one of the states into the other. This is repeated until each of the remaining states possess unique conditional future distributions. Such redundancies are fully classical in nature, and can readily be remedied by such classical means.

The next step is to consider \emph{why} the construction Eq.~\eqref{eqdeterministic} does not work for non-predictive generators. Simply replacing the predictive update rule with a sum over all possible end states weighted by their conditional probabilities will not yield the correct statistics. This is due to the lack of decoherence between different memory states in the outcome of the interaction: while the output register undergoes decoherence, superpositions of different end memory states given this output and the initial memory state are preserved; this leads to interference between the amplitudes of the their future morphs, in turn corrupting the statistics. To remedy this defect we must introduce a mechanism to break the coherence.

This can be achieved through the introduction of an auxiliary system that is imprinted with information about the end state into which the system transitions and subsequently discarded into the environment. The information to be encoded into this system has some freedom, and different encodings will yield different levels of memory compression. Here we will take a direct approach and encode the label of the end memory state; we discuss other possibilities later. The analogue of the interaction Eq.~\eqref{eqdeterministic} for our implementation becomes
\begin{equation}
\label{eqnondeterministic}
U\ket{\sigma_s}\ket{0}\ket{0}=\sum_{s'x}\sqrt{T_{s's}^x}\ket{\sigma_{s'}}\ket{x}\ket{s'},
\end{equation}
where we have set all $\{\varphi\}$ to zero. After discarding the auxiliary space and decohering the output space we obtain a channel of the form Eq.~\eqref{eqchanneldef}. We emphasise that while the information discarded into the auxilliary system is classical, the system itself must nevertheless be quantum, as it becomes entangled with the memory states. Intuitively, one can understand why the introduction of this system works by analogy with purification~\cite{nielsen2000quantum}; the evolution essentially creates the purified form of the target memory state distribution, such that when the auxilliary system is traced out the corresponding marginal in the memory system is as desired. That is, the auxilliary system provides a precise mechanism for breaking apart superpositions of memory states.

 We can express the channel $\Phi_{qg}$ in terms of Kraus operators~\cite{nielsen2000quantum} $\{L_s^x\}$:
\begin{equation}
\label{eqchannel}
\Phi_{qg}(\rho):=\sum_{s'x}L^x_{s'}\rho {L^x_{s'}}^\dagger,
\end{equation}
where $L_{s'}^x\ket{\sigma_s}:=\sqrt{T_{s's}^x}\ket{\sigma_{s'}}\ket{x}$. We can also break this down into a set of Kraus operators $\{K_s^x\}$ acting purely on the memory subspace, defined through $L_s^x:=K_s^x\otimes\ket{x}$. Armed with a viable quantum implementation of a non-deterministic HMM, we can now evaluate its memory compression advantage. We use $C_{qg}$ and $D_{qg}$ to represent the memory cost of the quantum implementation of a generator $g$, and similarly $C_{cg}$ and $D_{cg}$ for the classical implementation.

{\bf Theorem 1} (Memory compression advantage): \emph{The quantum implementation $(\{\ket{\sigma_s}\},\Phi_{qg})$ of Eq.~\eqref{eqchannel} achieves a memory compression advantage relative to the classical implementation $C_{qg}<C_{cg}$ whenever the generator $(\mathcal{S},\mathcal{X},\{T_{s's}^x\})$ is non-retrodictive.}

From Eq.~\eqref{eqnondeterministic}, we can use the unitarity of $U$ to find the overlaps of the quantum memory states:
\begin{align}
\label{eqoverlaps}
\braket{\sigma_j}{\sigma_{k}}&=\bopk{\sigma_j}{U^\dagger U}{\sigma_{k}}\nonumber\\
&=\sum_{j'k'xy}\sqrt{T_{j'j}^xT_{k'k}^y}\braket{\sigma_{j'}}{\sigma_{k'}}\braket{x}{y}\braket{j'}{k'}\nonumber\\
&=\sum_{j'x}\sqrt{T_{j'j}^xT_{j'k}^x}.
\end{align}
Thus, there is a non-zero overlap between any pair of quantum states iff both states have at least one possible outgoing transition for which the symbol and end state are identical. With these overlaps, we can then use the Gram matrix~\cite{horn1990matrix} of the stationary state to determine the memory costs of the implementation. The Gram matrix of the stationary state $\rho$ is given by \mbox{$\rho_G:=\sum_{ss'}\sqrt{\pi(s)\pi(s')}\braket{\sigma_s}{\sigma_{s'}}\ket{s}\bra{s'}$}, and possesses the same spectrum as $\rho$. Thus, by determining the spectrum of $\rho_G$ we are able to calculate the memory costs Eq.~\eqref{eqmeasures}. Further, it is evident that the diagonal elements of $\rho_G$ are given by $\pi(s)$, and so the classical stationary state is obtained from the Gram matrix by a projection in the classical memory state basis. Since projective measurements can never decrease any R\'enyi entropy~\footnote{This follows from the Schur-Horn theorem (that the spectrum of a Hermitian matrix majorises its diagonal elements~\cite{nielsen2001characterizing}) and the Schur-concavity of R\'{e}nyi entropies~\cite{marshall1979inequalities}.}, we can immediately conclude that $D_{qg}\leq D_{cg}$ and $C_{qg}\leq C_{cg}$: the quantum implementation never performs worse than the classical in terms of memory compression.

Moreover, for any R\'enyi entropy other than $\alpha=0$ (corresponding to $D_g$), projective measurements \emph{strictly} increase the entropy unless the projection leaves the state unchanged. Such a projection will preserve the state iff the off-diagonals in the projection basis are all zero. Thus, for the stationary states of our quantum implementations, there is a strict memory compression advantage $C_{qg}<C_{cg}$ iff there is at least one pair of states $(s,s'\neq s)\in\mathcal{S}^2$ for which $\braket{\sigma_s}{\sigma_{s'}}\neq0$. For no such pair to exist, there must be no two states in the generator that share a common transition in which the symbol and end state are the same. If there are no such states, then given the symbol and end state of a transition it is possible to specify the start state with certainty; this is precisely the definition of a retrodictive generator. Thus, there is no memory compression advantage to a quantum implementation of a generator iff it is retrodictive. 

From the overlaps Eq.~\eqref{eqoverlaps} it is possible to decompose the quantum memory states $\{\ket{\sigma_s}\}$ and $U$ into an orthogonal basis through a reverse Gram-Schmidt procedure~\cite{dennery1996mathematics}, analogous to methods in the case of deterministic generators~\cite{binder2018practical}.

\section{Thermal efficiency}

There is a deep physical connection between information theory and thermodynamics~\cite{leff2002maxwell}, first hinted at by Maxwell's demon~\cite{maxwell1872theory} and Szilard's engine~\cite{szilard1929entropieverminderung}, and captured more formally by Landauer's work on the dissipative costs of irreversible computation~\cite{landauer1961irreversibility}. Central to this is the information processing second law~\cite{deffner2013information, parrondo2015thermodynamics}, which -- mirroring its thermodynamical namesake -- places bounds on the entropic costs of changing a system's configuration. In the present context, the structured pattern produced by a generator outputting a stochastic process forms an information reservoir that can be harvested as a work source~\cite{mandal2012work, garner2017thermodynamics, boyd2017leveraging}. The work cost involved in producing the pattern depends on the generator and its implementation; we now show that our quantum implementations can also offer enhanced thermal efficiency. This complements and extends recent results on the thermodynamics of quantum implementations of predictive generators~\cite{loomis2020thermal, loomis2020thermodynamically}.

Fundamentally, the information processing second law mandates that the work cost for any generator to produce a pattern is bounded by the entropy of the pattern. That is, no generator can achieve a work cost $W<W_{\mathrm{min}}:=-\kb T \ln [2]h_\mu$, where $h_\mu:=\lim_{t\to\infty}H(X_{0:t})/t$ is the process' entropy rate. However, typical generators incur additional dissipative costs beyond this bound -- a so-called \emph{modularity} or \emph{locality dissipation} -- due to the generator acting in a time-local manner to produce the pattern~\cite{boyd2018thermodynamics}. An implementation of a generator is said to be \emph{thermally inefficient} if this additional cost is non-zero~\cite{loomis2020thermodynamically}.

In Appendix \ref{secappthm2} we show that results previously derived for the cost of quantum implementations of predictive generators~\cite{loomis2020thermal} can be directly applied to our quantum implementations of general generators. In particular, for an asymptotically-large ensemble of identical implementations of a generator $g$ acting independently in parallel, the average work cost per timestep per implementation is given by 
\begin{equation}
\label{eqthermocost}
W_{qg}=\kb T\ln[2](I(S';X)-H(X)),
\end{equation}
where $I(S';X)$ is the (quantum) mutual information between the two subspaces of the state \mbox{$\sum_{s,s',x}\pi(s)T_{s's}^x\ket{\sigma_{s'}}\bra{\sigma_{s'}}\otimes\ket{x}\bra{x}$}. An equivalent expression holds for the work cost $W_{cg}$ of the classical implementation. With this, we are able to establish a clear link between memory compression and thermal efficiency.

{\bf Theorem 2} (Thermal inefficiency allows quantum advantages): \emph{Iff the classical implementation of a generator $(\mathcal{S},\mathcal{X},\{T_{s's}^x\})$  is thermally inefficient, the quantum implementation $(\{\ket{\sigma_s}\},\Phi_{qg})$ of Eq.~\eqref{eqchannel} will both exhibit memory compression advantage $C_{qg}<C_{cg}$, and mitigate at least some of the thermal inefficiency $W_{qg}<W_{cg}$.}

For both the classical and quantum implementations the second term of Eq.~\eqref{eqthermocost} is identical, as it is a function of the process only. Thus, any thermodynamical advantage must be found by a reduction in the first term alone. The data processing inequality~\cite{cover2012elements} informs us that the mutual information between two variables is non-increasing under local transformations of one of the variables, and leaves it unchanged iff the transformation is reversible. The mutual information terms of the two implementations are linked by the mapping $\{\ket{s}\to\ket{\sigma_s}\}$ on the memory subsystem; crucially, this mapping is reversible iff the quantum memory states are all mutually orthogonal -- i.e., they are all non-overlapping. This is identical to the condition for when memory compression cannot be achieved. Thus, iff $C_{qg}=C_{cg}$ then $W_{qg}=W_{cg}$, and 
\begin{equation}
\label{eqmemwork}
C_{qg}<C_{cg}\Leftrightarrow W_{qg}<W_{cg}.
\end{equation}
Finally, it has previously been shown that a classical implementation has no thermal inefficiency iff the generator is retrodictive~\cite{boyd2018thermodynamics, garner2019oracular, loomis2020thermodynamically}. Since the quantum implementation of any non-retrodictive generator achieves memory compression, they must also too reduce the thermal inefficiency, by Eq.~\eqref{eqmemwork}.

It is natural to then ask whether quantum implementations can bypass the locality dissipation entirely, and if so, under what conditions. The answer provides a complementary view to the previous theorem.

{\bf Theorem 3} (Non-eradication of dissipation): \emph{A quantum implementation $(\{\ket{\sigma_s}\},\Phi_{qg})$ of a generator $(\mathcal{S},\mathcal{X},\{T_{s's}^x\})$ that operates with no locality dissipation can be simulated by a classical implementation at no additional memory cost. Any reduction of thermal inefficiency by a quantum implementation is only mitigation, not eradication; any thermally efficient quantum implementation yields no memory compression advantage.}

The proof is given in Appendix \ref{secappthm3}; it utilises a framework developed in proving a similar statement for quantum implementations of predictive generators~\cite{loomis2020thermodynamically}, effectively condensing and generalising this earlier derivation. The crux of this generalised theorem is that \emph{no} quantum implementation of \emph{any} generator can simultaneously achieve an advantage in memory compression and perfect thermal efficiency. This highlights that optimising efficiency in the modelling of stochastic processes involves a trade-off: generators with minimally-dissipative classical implementations are often not the most memory-efficient -- this frustration extends into the quantum domain.

\section{Example: Simple Nonunifilar Source}

\begin{figure}
\includegraphics[width=\linewidth]{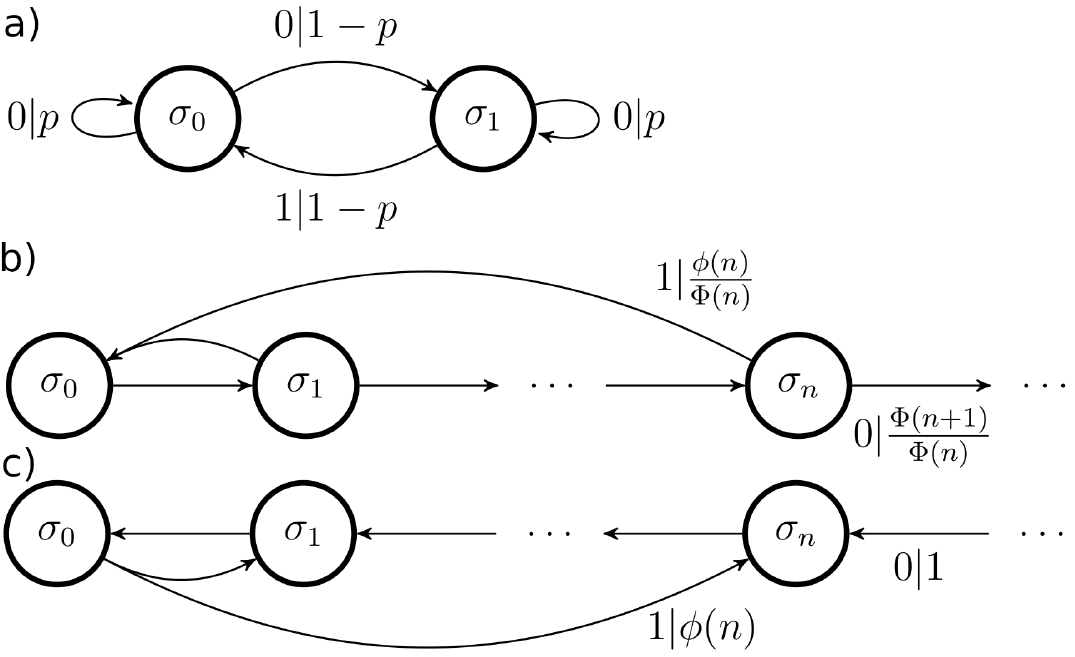}
\caption{Different possible HMM generators of the Simple Nonunifilar Source process. a) Two state nondeterministic generator, b) minimal memory predictive generator, and c) minimal retrodictive generator. Notation $x|p$ denotes that with probability $p$ the indicated transition occurs accompanied by the emission of symbol $x$.}
\label{fig}
\end{figure}

As an example, we consider the so-called Simple Nonunifilar Source (SNS) process, which allows us to compare and contrast memory and work costs of classical and quantum implementation deterministic and nondeterministic generators, and explore some of these trade-offs between thermal efficiency and compression. Depicted in \figref{fig}, we will look at (A) a two state nondeterministic generator of the process, (B) the minimal memory predictive generator (i.e., the $\varepsilon$-machine, and (C) the minimal retrodictive generator. As a measure of memory, we will consider the Shannon/von Neumann entropy of the steady-state, which we denote by $C_{MG}$, where $G=\{\mathrm{A,B,C}\}$ indicates which generator the measure is with respect to, and $M=\{c,q\}$ whether the implementation is classical or quantum. We will also consider the dimension of the implementations' memories, and the work costs $W_{MG}$. We will predominantly focus on the case where $p=1/2$, and later remark on the general case.

For the two state generator A, it is clear from inspection that the two memory states have equal steady-state occupation probabilities. Thus, we can immediately conclude $C_{c\mathrm{A}}=1$. For the quantum implementation, we have
\begin{align}
U\ket{\sigma_0}\ket{0}\ket{0}&=\frac{1}{\sqrt{2}}\ket{\sigma_0}\ket{0}\ket{0}+\frac{1}{\sqrt{2}}\ket{\sigma_1}\ket{0}\ket{1}\nonumber\\
U\ket{\sigma_1}\ket{0}\ket{0}&=\frac{1}{\sqrt{2}}\ket{\sigma_0}\ket{1}\ket{0}+\frac{1}{\sqrt{2}}\ket{\sigma_1}\ket{0}\ket{1}.\nonumber\\
\end{align}
We can use that $U^\dagger U=1$ to determine that $\braket{\sigma_0}{\sigma_1}=1/2$. Without loss of generality we can thence assign $\ket{\sigma_0}=\ket{0}$ and $\ket{\sigma_1}=(1/2)\ket{0}+(\sqrt{3}/2)\ket{1}$. The steady-state of the memory is given by $(1/2)(\ket{\sigma_0}\bra{\sigma_0}+\ket{\sigma_1}\bra{\sigma_1})$, from which we determine $C_{q\mathrm{A}}=0.811$, yielding a quantum compression advantage. Note that here and throughout the example all numerical values are given to three significant figures.

From Eq.~\eqref{seqimpcost}, the work cost is given by $W=H(S')-H(S'X)$, where $X$ is the emitted symbol, $S'$ the state of the system after emission, and we have normalised $\kb T\ln[2]=1$. The first term corresponds to the entropic memory $C$, and so we need only calculate the entropy of the combined final state of a transition with the output. For generator A, it can be straightforwardly deduced that $P(0,\sigma_0)=1/4$, $P(1,\sigma_0)=1/4$, and $P(0,\sigma_1)=1/2$. This allows us to calculate that $W_{c\mathrm{A}}=-0.5$ and $W_{q\mathrm{A}}=-0.558$, thus also yielding a quantum thermal advantage. Note that the work costs are negative as we are transforming a blank tape containing no information to a stochastic pattern with non-zero entropy.

Next, we look at the minimal memory predictive generator B. Let us first remark on the drastic difference in the number of states in the model -- an infinite number -- compared to the two of the nondeterministic generator A. This emphasises the potential power of turning to nondeterministic generators generally for compressing the memory dimension. To aid us in calculating the memory and work costs of generator B, we note that the SNS is a renewal process, allowing results from Refs.~\cite{marzen2015informational, elliott2018superior} to be put to use. Let us introduce some quantities from the modelling of renewal processes. The wait-time distribution $\phi(n)$ is defined as the probability that after a 1 is seen, there are $n$ consecutive 0s before the next 1. The survival probability $\Phi(n):=\sum_{n'=n}^\infty\phi(n')$ describes the probability that the number of 0s between a consecutive pair of 1s is at least $n$. Finally, the mean firing rate is defined as $\mu:=1/\sum_{n=0}^\infty\Phi(n)$. In previous works, it has been established that the steady-state probability distribution of the states $\sigma_n$ is given by $P(\sigma_n)=\mu\Phi(n)$, and that the states of the corresponding quantum implementation can be encoded as
\begin{equation}
\ket{\sigma_n}=\sum_{n'=0}^\infty\frac{\sqrt{\phi(n'+n)}}{\sqrt{\Phi(n)}}\ket{n'}.\nonumber
\end{equation}
Moreover, it has been shown that for the SNS, $\phi(n)=np^{n-1}(1-p)^2.$

With all this at hand, we can calculate $C_{c\mathrm{B}}=2.71$ and $C_{q\mathrm{B}}=0.386$. We observe that this quantum implementation has a lower entropic memory cost than that of generator A, exemplifying that different quantum implementations would be favoured depending on whether the dimension or the entropic cost of memory is prioritised~\cite{liu2019optimal}. To calculate the work cost, we use that $P(1,\sigma_0)=\sum_nP(\sigma_n)P(1,\sigma_0|\sigma_n)=\mu\sum_n\phi(n)=\mu$ and $P(0,\sigma_n)=P(\sigma_{n-1})P(0|\sigma_{n-1})=\mu\Phi(n)$ for $n=1:\infty$. Then, $W_{c\mathrm{B}}=0$ and $W_{q\mathrm{B}}=-0.468$. Thus, while the quantum implementation of B has a lower entropic cost than the implementations of A, it has a higher work cost than both -- highlighting that though quantum compression and thermal advantages go hand-in-hand relative to a classical implementation of the same generator, there is no such definitive heirarchy between quantum implementations of different generators.

Finally, for completeness we consider the retrodictive generator C, which should exhibit no locality dissipation. As the SNS is a renewal process, the classical implementation of generator C bears the same entropic memory cost as that of the minimal deterministic generator B~\cite{marzen2015informational}. Moreover, Theorem 1 tells us that there is no quantum compression advantage for a retrodictive generator. Thus, $C_{c\mathrm{C}}=C_{q\mathrm{C}}=2.71$. From Theorems 2 and 3, we can also deduce $W_{q\mathrm{C}}=W_{c\mathrm{C}}$; we now only need to calculate this latter quantity. We have that $P(1,\sigma_n)=P(\sigma_n|1)P(1)=\mu\phi(n)$ and $P(0,\sigma_n)=P(\sigma_{n+1})=\mu\Phi(n+1)$. Thus, we find $W_{c\mathrm{C}}=W_{q\mathrm{C}}=-0.678$. We can verify that this saturates the bound set by the information processing second law by checking that it is equal to the (negative of the) entropy rate of the process. This can be calculated from the deterministic generator B through $h_\mu=\sum_nP(\sigma_n)H(P(X|\sigma_n))$, which indeed yields $h_\mu=0.678$. Thus, we see that this example also highlights the general trade-off between compression and work cost, as the implementations with the lowest work cost are not those that achieve the best compression, by either measure of memory. That is, the efficient retrodictive generator C requires an infinite number of states and has a higher entropic cost than generator A, which may be compressed down to two dimensions, at the cost of some additional dissipation.

We summarise these results in Table \ref{table}.
\begin{table}[h!]
\begin{center}
\begin{tabular}{c|cc}
  & $c$     & $q$ \\ \hline
A & $C_{c\mathrm{A}}=1$, $W_{c\mathrm{A}}=-0.5$ & $C_{q\mathrm{A}}=0.811$, $W_{q\mathrm{A}}=-0.558$ \\
B & $C_{c\mathrm{B}}=2.71$, $W_{c\mathrm{B}}=0$ & $C_{q\mathrm{B}}=0.386$, $W_{q\mathrm{B}}=-0.468$ \\
C & $C_{c\mathrm{C}}=2.71$, $W_{c\mathrm{C}}=-0.678$ & $C_{q\mathrm{C}}=2.71$, $W_{q\mathrm{C}}=-0.678$
\end{tabular}
\caption{Summary of memory and thermal costs of classical and quantum implementations of different generators of the simple nonunifilar source process.}
\label{table}
\end{center}
\end{table}

Finally, let us discuss the case where $p\neq 1/2$. By varying $p$, the trade-off between thermal efficiency and compression can become even more marked. As $p\to1$, the entropic cost of implementing the thermally efficient generator $C_{c\mathrm{C}}$ (and $C_{q\mathrm{C}}$)  will diverge. Meanwhile, both the classical and quantum implementations of A have entropic costs bounded from above by 1. Indeed, a more detailed calculation finds that $C_{q\mathrm{A}}\to1$. Moreover, for all generators and implementations considered, it can be seen that the work costs all tend to 0 in this limit. That is, as $p\to1$, the increased compression offered by the two state nondeterministic generator implementations grows while their thermal inefficiency diminishes.

\section{Discussion}

Our work extends a series of results on the power and limitations of using quantum technologies to implement generators of stochastic processes by generalising to a much broader class of HMMs. At the core of this, we have shown that if a classical generator of a stochastic process has any thermal inefficiency due to its time-local operative nature, then it can be implemented with greater thermal and memory efficiency by a quantum device. We also provided a systematic method for determining the architecture of such an implementation. When perfect thermal efficiency is required however, the best classical and quantum implementations are one and the same. 

To achieve the necessary decoherence required to break apart superpositions of memory states for non-predictive generators, we introduced an additional auxiliary system that carries information about the transition. To provide a universal, systematic protocol we set this information to be the end state of the transition. This choice is not unique; for any transition $s$ to $s'$ with emitted symbol $x$, any state $\ket{\psi(s,s',x)}$ will suffice provided
\begin{equation}
\label{eqjunkcond}
\braket{\psi(s,s',x)}{\psi(s,s'',x)}=\delta_{s's''}
\end{equation}
for all pairs of possible transitions. The choice of $\{\ket{\psi(s,s',x)}\}$ affects the overlaps of the quantum memory states through Eq.~\eqref{eqoverlaps}, potentially allowing for further thermal and memory advantages. Particularly, for generators of highly non-Markovian processes in which most transitions have only a small number of possible end states given the initial state and symbol, setting $\ket{\psi(s,s',x)}=\ket{s'}$ will likely be rather inefficient~\cite{mahoney2016occam,binder2018practical,elliott2020extreme}. We leave the optimisation of this freedom to future work. The use of complex phases to improve memory efficiency in quantum implementations of predictive generators~\cite{liu2019optimal} is a special case of this auxiliary space, restricted to one dimension. The theorems above hold for any choice of $\{\ket{\psi(s,s',x)}\}$, provided that they satisfy Eq.~\eqref{eqjunkcond} and do not remove all possible non-orthogonalities that can be engineered between quantum memory states. We also remark that the structure of this additional system corroborates the interpretation that non-predictive generators contain `sideband' information about the future of the process they generate~\cite{garner2019oracular}: were an external party to retrieve this system from the environment after it is discarded, they would be able to better anticipate the future of the process than they could from observing its past alone.

As with quantum implementations of predictive generators, proof-of-principle demonstrations of these advantages are within reach of current experiments~\cite{palsson2017experimentally, ghafari2019dimensional, ghafari2019interfering}. In taking quantum memory and thermodynamical advantages in stochastic modelling to a more general medium, we open up a number of future research avenues, some paralleling developments in quantum implementations of predictive generators, some unique to the wider spectrum. For example, our results can be further extended to encompass input-output~\cite{barnett2015computational, thompson2017using} and continuous-time processes~\cite{marzen2015informational, marzen2017informational, marzen2017structure, elliott2018superior, elliott2019memory, elliott2020extreme, elliott2021quantum} as well as inference protocols~\cite{ho2020robust}, and the trade-off between memory- and thermal-efficiency can be explored to determine the generator that best compromises the two -- and whether classical and quantum implementations agree on which generator this should be. This latter direction will also require developments even in the purely classical setting: while the optimal predictive generators can be systematically found, this remains an open question across all generators in general. Nevertheless, our results show that no matter how good a classical implementation is, a quantum counterpart is almost certainly better.

\acknowledgements
We thank Andrew Garner, Varun Narasimhachar and Mile Gu for discussions. This work was funded by the Imperial College Borland Fellowship in Mathematics and grant FQXi-RFP-1809 from the Foundational Questions Institute and Fetzer Franklin Fund (a donor advised fund of the Silicon Valley Community Foundation).

\appendix

\section{Work cost for quantum implementations}
\label{secappthm2}

In Ref.~\cite{loomis2020thermal} it was shown that for $N$ independent copies of a system $R$ and blank ancilla undergoing unitary evolutions to correlated states of the joint system-ancillas $R'A$ there is a procedure of erasing the $A$ to obtain $R'$ alone that is successful with probability $1-2^{-\sqrt{N}}$, bearing a work cost per copy $W$ of
\begin{equation}
\label{seqworkcost}
\frac{W}{\kb T\ln[2]}= H(R)-H(R')
\end{equation}
as $N\to\infty$, where $H(.)$ is the von Neumann entropy~\cite{nielsen2000quantum}. We do not reproduce this derivation here, and refer the interested reader to Ref.~\cite{loomis2020thermal}. We will however show how it can be deployed in our context.

Recall from the main text that we have the evolution
\begin{equation}
\label{seqnondeterministic}
U\ket{\sigma_s}\ket{0}\ket{0}=\sum_{s'x}\sqrt{T_{s's}^x}\ket{\sigma_{s'}}\ket{x}\ket{s'}.
\end{equation}
With a further unitary step that copies the output onto another auxiliary space such that we obtain a final state
\begin{equation}
\sum_{s'x}\sqrt{T_{s's}^x}\ket{\sigma_{s'}}\ket{x}\ket{s'}\ket{x},
\end{equation}
we can then trace out the two auxiliary spaces to obtain the desired evolution according to Eq.~\eqref{eqchanneldef}. In terms of Eq.~\eqref{seqworkcost} the initial system state corresponds to the initial memory state $S$, the final system state is the combined final memory state and output $S'X$, and the ancillae to be reset are the two auxiliary spaces. Thus, the work cost for our evolution is given by
\begin{equation}
\label{seqimpcost}
\frac{W_{qg}}{\kb T\ln[2]}= H(S)-H(S'X),
\end{equation}
where here for shorthand we use $S$ and $S'$ to represent the states corresponding to the intial and final memory states. Noting that $H(S)=H(S')$ for a memory initialised in the stationary state of the evolution, we can recast this as
\begin{equation}
W_{qg}=\kb T\ln[2] (I(S';X)-H(X)),
\end{equation}
as given in the main text. This matches the result for quantum implementations of deterministic generators as the principal difference is the extra auxiliary space added to the ancilla -- which does not appear directly in the work cost Eq.~\eqref{seqworkcost}. Moreover, by including a further auxiliary space as part of the ancilla that after the evolution also outputs the initial memory state we recover the classical implementation where all memory states are orthogonal -- and therefore see that Eq.~\eqref{seqimpcost} similarly holds classically. Note that these work costs are negative; positive work can be extracted by the generator implementations~\cite{loomis2020thermal}

\section{Proof of Theorem 3}
\label{secappthm3}

In Ref.~\cite{loomis2020thermodynamically} it was shown that a quantum implementation of a deterministic generator can operate with zero modularity cost only if the generator is also retrodictive, and hence does not allow for any quantum memory advantage. Theorem 3 declares that this result holds for quantum implementations in general. Our proof mirrors many aspects of that for the deterministic case~\cite{loomis2020thermodynamically}, with appropriate generalisation. Before proceeding with our proof, we introduce a definition and result involved in the proof of the deterministic case.

{\bf Definition 2} (Maximal local commuting measurement)~\cite{loomis2020thermodynamically}: \emph{Given a bipartite state $\rho_{AB}$, a} Maximal local commuting measurement \emph{(MLCM) of $A$ for $B$ is a local measurement $X$ on $A$ such that}
\begin{equation}
\rho_{AB}=\bigoplus_xP(x)\rho_{AB}^x
\end{equation}
\emph{where $\rho_{AB}^x:=(\Pi_X^x\otimes I_B)\rho_{AB}(\Pi_X^x\otimes I_B)$ and $\Pi_X^x$ are the projection operators corresponding to the measurement, and no further non-trivial local measurement $Y$ on $A$ can be performed without disturbing $\{\rho_{AB}^x\}$.}

The MLCM is proven to be unique~\cite{loomis2020thermodynamically}.

{\bf Proposition 1} (Reversible local operations)~\cite{loomis2020thermodynamically}: \emph{Given a bipartite state $\rho_{AB}$ and a local operation $\Phi_A$ on $A$ such that $\rho_{CB}:=(\Phi_A\otimes I_B)[\rho_{AB}]$, let $X$ be the MLCM of $A$ for $B$, and $Y$ the MLCM of $C$ for $B$. Then, $I(A;B)=I(C;B)$ iff $\Phi_A$ can be expressed by Kraus operators $\{K^\alpha\}$ of the form}
\begin{equation}
K^\alpha=\bigoplus_{xy}e^{i\varphi_{xy\alpha}}\sqrt{P(y,\alpha|x)}U^{y|x},
\end{equation}
\emph{where $\{\varphi_{xy\alpha}\in\mathbb{R}\}$ are arbitrary and $P(Y,\alpha|X)$ is a stochastic channel that is non-zero only when $\rho_{AB}^x$ and $\rho_{CB}^y$ are equivalent up to a local unitary operation $U^{y|x}$.}

The proof is given in Ref.~\cite{loomis2020thermodynamically}.

We are now in a position to prove Theorem 3, repeated here for convenience.

{\bf Theorem 3} (Non-eradication of dissipation): \emph{A quantum implementation $(\{\ket{\sigma_s}\},\Phi_{qg})$ of a generator $(\mathcal{S},\mathcal{X},\{T_{s's}^x\})$ operates with no locality dissipation iff it can be simulated by a classical implementation at no additional memory cost. Any reduction of thermal inefficiency by a quantum implementation is only mitigation, not eradication; any thermally efficient quantum implementation yields no memory compression advantage.}

Define $\rho_g(t)$ be the combined state of the system $X_{0:t}S_t$ formed from the memory state and previous $t$ outputs. Let $\Theta$ be the MLCM of $S_t$ for $X_{0:t}$ such that $\rho_g(t)=\bigoplus_\theta P(\theta)\rho_g^\theta(t)$ where $\rho_g^\theta(t):=(I_{X_{0:t}}\otimes\Pi^\theta)\rho_g(t)(I_{X_{0:t}}\otimes\Pi^\theta)$, and similarly let $\Theta'$ be the MLCM of $S_{t+1}$ for $X_{0:t+1}$. 

Recall from Eq.~\eqref{eqchannel} that we have
\begin{equation}
\label{seqchannel}
\Phi_{qg}(\rho):=\sum_{s'x}L^x_{s'}\rho {L^x_{s'}}^\dagger,
\end{equation}
where we can further decompose $L_s^x:=K_s^x\otimes\ket{x}$ to obtain Kraus operators $\{K_s^x\}$ acting on the memory space alone. From Proposition 1 if there is to be no locality dissipation we require these to take the form
\begin{equation}
\label{eqnoloccond}
K^x_{s'}=\bigoplus_{\theta\theta'}\sqrt{P(s',x,\theta'|\theta)}U^{\theta'x|\theta}.
\end{equation}

Consider now that a further unitary operation is used to imprint the label of the end state of the transition as an additional subspace of the output pattern. The implementation is now of the generator $(\{\theta\},\mathcal{S}\mathcal{X},\{T_{\theta'\theta}^{sx}\})$ that produces a joint stochastic process combining the outputs of the original stochastic process, and the trajectory of future memory states. Crucially though, this shares the same memory cost as the implementation of the original generator, and has the same form for the $\{K_s^x\}$ as Eq.~\eqref{eqnoloccond}. However, for Eq.~\eqref{eqnoloccond} to hold, the new generator must be retrodictive, i.e., given $(x,s',\theta')$, $\theta$ is uniquely determined~\cite{loomis2020thermodynamically}. Since the generator is retrodictive, the quantum implementation achieves no memory compression relative to a classical implementation of the same generator. Thence, there exists a classical implementation of the retrodictive generator of the joint process that bears the same memory cost as the quantum implementation of the original generator. This classical implementation can then be used to generate the process associated with the original generator by coarse-graining its output to discard the parts associated with the memory state trajectory.

Thus, whenever a quantum implementation of a generator achieves no locality dissipation, a classical implementation of the same process can be constructed that has the same memory cost. This yields the content of Theorem 3.

\bibliography{ref}

\end{document}